\documentclass[twocolumn,showpacs,preprintnumbers,amsmath,amssymb, superscriptaddress, prb]{revtex4}

\usepackage{graphicx}
\usepackage{dcolumn}
\usepackage{bm}

\usepackage{color}
\definecolor{red}{rgb}{1,0,0}

\definecolor{blue}{rgb}{0,0,1}

\definecolor{green}{rgb}{0,1,0}

\begin{document}
\preprint{APS}

\author {Mariano de Souza}
\email{mariano@rc.unesp.br} \affiliation{Physikalisches Institut,
Goethe-Universit\"{a}t Frankfurt,  SFB-TR49, Max-von-Laue Strasse,
1, D-60438 Frankfurt am Main, Germany}\affiliation{Departamento de
F\'isica, IGCE, Unesp - Universidade Estadual Paulista, Cx.\,Postal
178, 13500-970 Rio Claro (SP), Brazil}
\author{Andreas Br\"uhl}
\affiliation{Physikalisches Institut,  Goethe-Universit\"{a}t
Frankfurt,  SFB-TR49, Max-von-Laue Strasse, 1, D-60438 Frankfurt am
Main, Germany}
\author{Christian Strack}
\affiliation{Physikalisches Institut,  Goethe-Universit\"{a}t
Frankfurt,  SFB-TR49, Max-von-Laue Strasse, 1, D-60438 Frankfurt am
Main, Germany}
\author{Dieter Schweitzer}
\affiliation{3.\,Physikalisches Institut, Universit\"{a}t Stuttgart,
D-70550 Stuttgart, Germany}
\author{Michael Lang}
\affiliation{Physikalisches Institut,  Goethe-Universit\"{a}t
Frankfurt,  SFB-TR49, Max-von-Laue Strasse, 1, D-60438 Frankfurt am
Main, Germany}




\title{Magnetic Field-Induced Lattice Effects in a Quasi-2D Organic Conductor
Close to the Mott Metal-Insulator Transition}
%

\begin{abstract}
We present ultra-high-resolution dilatometric studies in magnetic
fields on a quasi-two-dimensional organic conductor $\kappa$-(D8-BEDT-TTF)$_{2}$Cu[N(CN)$_{2}$]Br, which is
located close to the Mott metal-insulator (MI) transition. The
obtained thermal expansion coefficient, $\alpha(T)$, reveals two
remarkable features: (i) the Mott MI transition temperature $T_{MI}$
= (13.6 $\pm$ 0.6)\,K is insensitive to fields up to 10\,T, the
highest applied field; (ii) for fields along the interlayer
\emph{b}-axis, a magnetic-field-induced (FI) phase transition at
$T_{FI}$ = (9.5 $\pm$ 0.5)\,K is observed above a threshold field
$H_c \sim$ 1 T, indicative of a spin reorientation with strong
magneto-elastic coupling.
\end{abstract}


\maketitle


\date{\today}


\section{Introduction}
\begin{figure}
\begin{center}
\includegraphics[angle=0,width=0.45\textwidth]{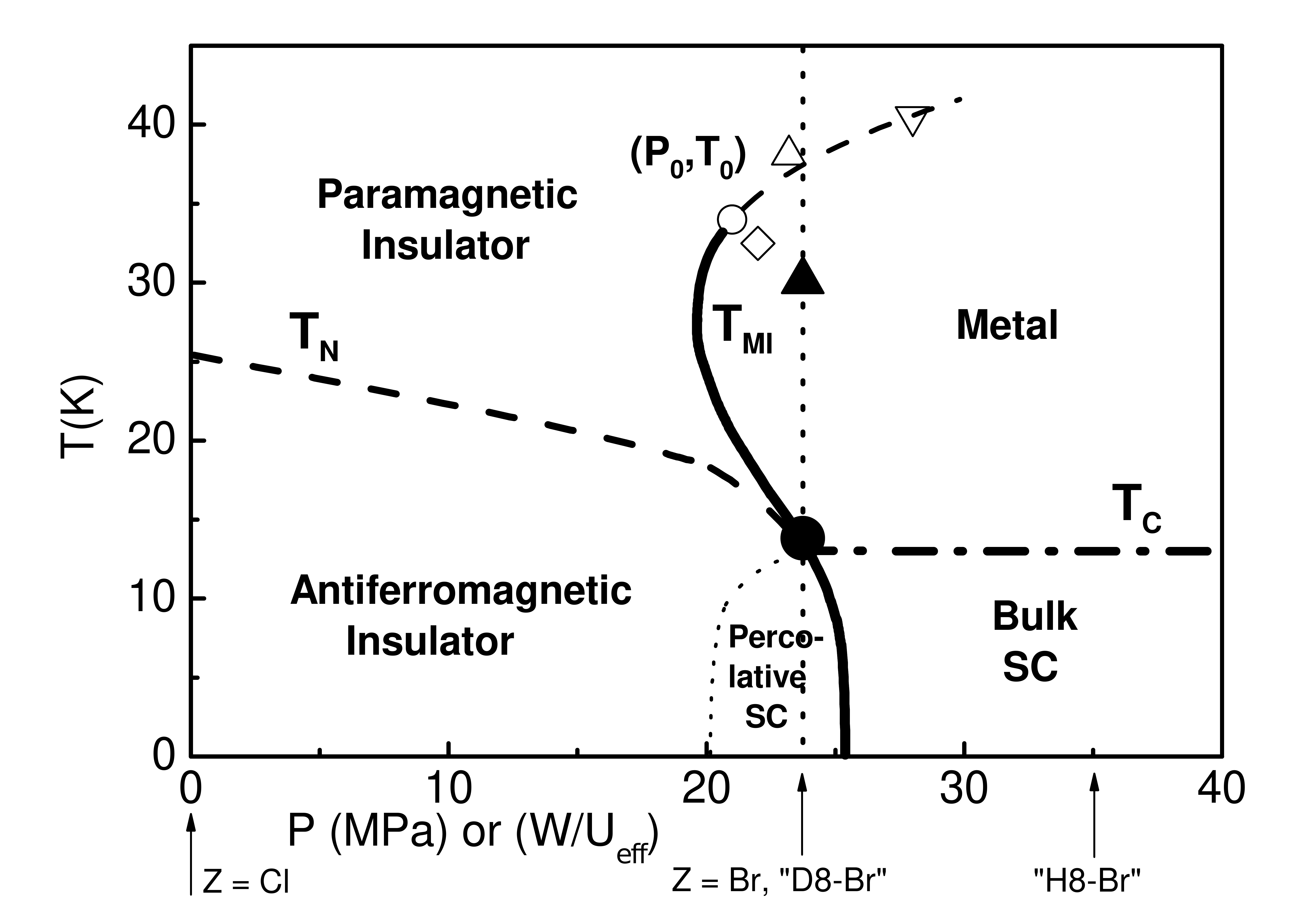}
\end{center}
\caption{Pressure-temperature, $\emph{P}-\emph{T}$, phase diagram of
$\kappa$-(ET)$_{2}$Cu[N(CN)$_{2}$]Z at zero magnetic field. Broken line labeled $T_N$ indicates the paramagnetic to antiferromagnetic transition, the broken-dotted line labeled $T_c$ corresponds to the transition into bulk superconductivity and the thick solid line labeled $T_{MI}$ marks the first-order metal-to-insulator transition. D8-Br and H8-Br
(position estimated according to \cite{LangM2S}) refer to the
position of the deuterated and protonated single crystals of
$\kappa$-(ET)$_{2}$Cu[N(CN)$_2$]Br, respectively. Vertical dotted
line indicates the performed $T$-sweeps for "D8-Br" crystals, illustrating a crossing of the
\emph{S}-shaped line. Open symbols refer to literature data for the
critical end point ($P_0$, $T_0$): ($\diamond$) \cite{Lefebvre 00},
($\bigtriangledown$) \cite{Limelette 03}, ($\bigcirc$)
\cite{Fournier 03}, {$\bigtriangleup$} \cite{Kagawa 04},
$\blacktriangle$ \cite{Mariano}. Position of full circle, corresponding to $T_{MI}$ for the present crystal, implies that at this point in the phase diagram $T_{MI}$ coincides with $T_N$, cf. \cite{LangSCES07}.}\label{Fig.1}
\end{figure}

Currently, strong activities in condensed matter physics have been
directed towards a better understanding of correlation effects
in low-dimensional systems. The strong interaction between electrons
in these systems gives rise to several interesting phenomena. Among
them, the Mott metal-to-insulator (MI) transition can be considered
as one of the most prominent examples. Organic conductors of the
$\kappa$-phase (BEDT-TTF)$_{2}$X family (BEDT-TTF, or simply ET,
refers to the donor molecule bis(ethylenedithio)tetrathiafulvalene
and X to a monovalent anion) have been recognized as appropriate
systems for studying phenomena originating from the interplay of
electron-electron and electron-lattice interactions in reduced
dimensions, for a recent review see, e.g. \cite{Lang}. These
substances are built by layers of interacting dimers,
i.e.\,(ET)$_2^+$, sandwiched by sheets of insulating polymeric
counter anions X$^{-}$, giving rise thus to a quasi-two-dimensional
electronic structure. Their ground states can be tuned by chemical
substitution and/or applying external pressure, see, e.g., the
pressure-temperature ($P$-$T$) phase diagram in \,Fig.\,\ref{Fig.1}.
While the salt with X = Cu[N(CN)$_{2}$]Cl ($\kappa$-Cl in short)
is an antiferromagnetic Mott insulator (AFI) with $T_N$ $\approx$
26\,K, the salt with X = Cu[N(CN)$_{2}$]Br ($\kappa$-H8-Br in short)
is a superconductor (SC) with $T_c$ $\approx$ 11\,K, the highest
$T_c$ at ambient pressure among all ET-based organic compounds
investigated up to date.
Interestingly enough, for X = Cu[N(CN)$_{2}$]Br, the exchange of the
hydrogen atoms of the ethylene end groups of the ET molecules by
deuterium results in an antiferromagnetic insulating ground state,
corresponding to a shift on the pressure axis towards lower pressure
\cite{Kawamoto97}. The insulating state in the $P$-$T$ phase diagram
is separated from the metallic/superconducting range by an
\emph{S}-shaped first-order phase transition line (thick solid line
in Fig.\,\ref{Fig.1}) which ends in a second-order critical point,
indicated by ($P_0$, $T_0$) in Fig.\,\ref{Fig.1}, see
Ref.\cite{Mariano} and references therein for details. Hence, due to
their close proximity to the MI boundary, fully deuterated salts of
$\kappa$-(ET)${_2}$Cu[N(CN)$_{2}$]Br, abbreviated to $\kappa$-D8-Br
hereafter, have been recognized as suitable systems for exploring
the Mott MI transition. The magnetic properties of the
$\kappa$-phase (ET)$_{2}$X family, especially the X =
Cu[N(CN)$_{2}$]Cl salt, have been investigated by various
experimental techniques, see, e.g. refs.\cite{Welp,
Pinteric,Miyagawa}. From the analysis of the NMR line shape,
relaxation rate and magnetization data, K. Miyagawa \emph{et al.}
\cite{Miyagawa} were able to describe the spin structure of this
state. Below $T_N$ = 26 -- 27\,K, they found a commensurate
antiferromagnetic ordering with a magnetic moment of (0.4 --
1.0)$\mu_B$/dimer. The observation of an abrupt jump in the
magnetization curves for magnetic fields applied perpendicular to
the conducting layers, i.e.\,along the \emph{b}-axis, was attributed
to a \emph{spin-flop} (SF) transition. Furthermore, a detailed
discussion about the spin-reorientation, taking into account the
Dzialoshinskii-Moriya interaction, was presented by D.\,F.\,Smith
\emph{et al.} \cite{Dylan1, Dylan2}. By resistance measurements
under control of temperature, pressure and magnetic field, a
magnetic field-induced Mott MI transition was observed by F. Kagawa
\emph{et al.} \cite{Kagawa 04a}. Interestingly enough, similar to
this finding for the pressurized $\kappa$-Cl salt, a magnetic
field-induced MI transition was also observed in partially
deuterated $\kappa$-(ET)$_{2}$Cu[N(CN)$_{2}$]Br \cite{Kawamoto}.
Furthermore, from $^{12}$C-NMR studies evidence for phase separation
into metallic/superconducting and magnetic phases in $\kappa$-D8-Br
was reported in the literature \cite{Miyagawa1}. From
temperature-dependent measurements of the coefficient of thermal
expansion, the role of the lattice degrees of freedom for the Mott
transition \cite{Mariano} and the Mott criticality \cite{Bartosch}
were studied. The latter results were found to be at odds with the
Mott criticality derived from conductivity \cite{Kagawa05} and NMR
\cite{Kagawa09} studies on pressurized $\kappa$-Cl. In order to gain
more insight into the nature of the state on the insulating side of
the Mott transition for the present $\kappa$-D8-Br material, we have
performed thermal expansion measurements in magnetic fields. In this
communication, we present expansivity data on single crystalline
$\kappa$-D8-Br in magnetic fields up to 10\,T and explore the field
effects on the various phases in the vicinity of the MI line.

\section{Experimental}
High-quality single crystals of fully deuterated
$\kappa$-(D8-ET)$_{2}$Cu[N(CN)$_{2}$]Br were prepared according to an
alternative procedure as described in Ref.\cite{Crystals1, Crystals2}.
The single crystal studied here, labelled as \#3 (batch A2907), is
identical to the one studied in Ref.\cite{LangM2S}. The linear thermal
expansion coefficient, $\alpha(\textit{T})=\textit{l}^{-1}(\partial
\textit{l}/\partial \textit{T})$ ($l$ is the sample length), was measured by employing an
ultra-high-resolution capacitance dilatometer with a maximum
resolution of $\Delta l/l=10^{-10}$, built after \cite{Pott20a}.
Samples of the organic charge-transfer salts studied here are very
sensitive to the quasi-uniaxial pressure exerted by the dilatometer
\cite{PhD}. In fact, the uniaxial pressure acting on the crystal, typically a few
bars, can be adjusted by setting the starting capacitance. In order
to reduce the strain exerted by the dilatometer on the sample to a
minimum, a very small starting capacitance was chosen.
The experimental data presented were corrected only for the thermal
expansion of the dilatometer cell with no further data processing.
The alignment of the crystal was guaranteed with an error margin of
$\pm$3$^o$. In all measurements, the magnetic field is parallel to
the measuring direction. Resistance measurements were carried out by
employing the standard four-terminal ac technique. In order to
reduce cooling-rate-dependent effects associated with disorder of
the ethylene end groups of the ET molecules, a cooling rate of
$\sim$$-$3\,K/h (thermal expansion) and $\sim$$-$6\,K/h (resistance)
through the glasslike transition around 77\,K \cite {Mueller 02} was
applied. After the initial controlled cool-down, the sample was kept
at temperatures below 40\,K. Measurements of $\alpha(T, B = const.)$
were performed at temperatures ranging from 4.5\,K up to about 14\,K
except for $B$ = 0.5\,T where the measurements were limited to $T
\leq$ 11\,K.

\section{Results and Discussion}
\begin{figure}
\begin{center}
\includegraphics[angle=0,width=0.42\textwidth]{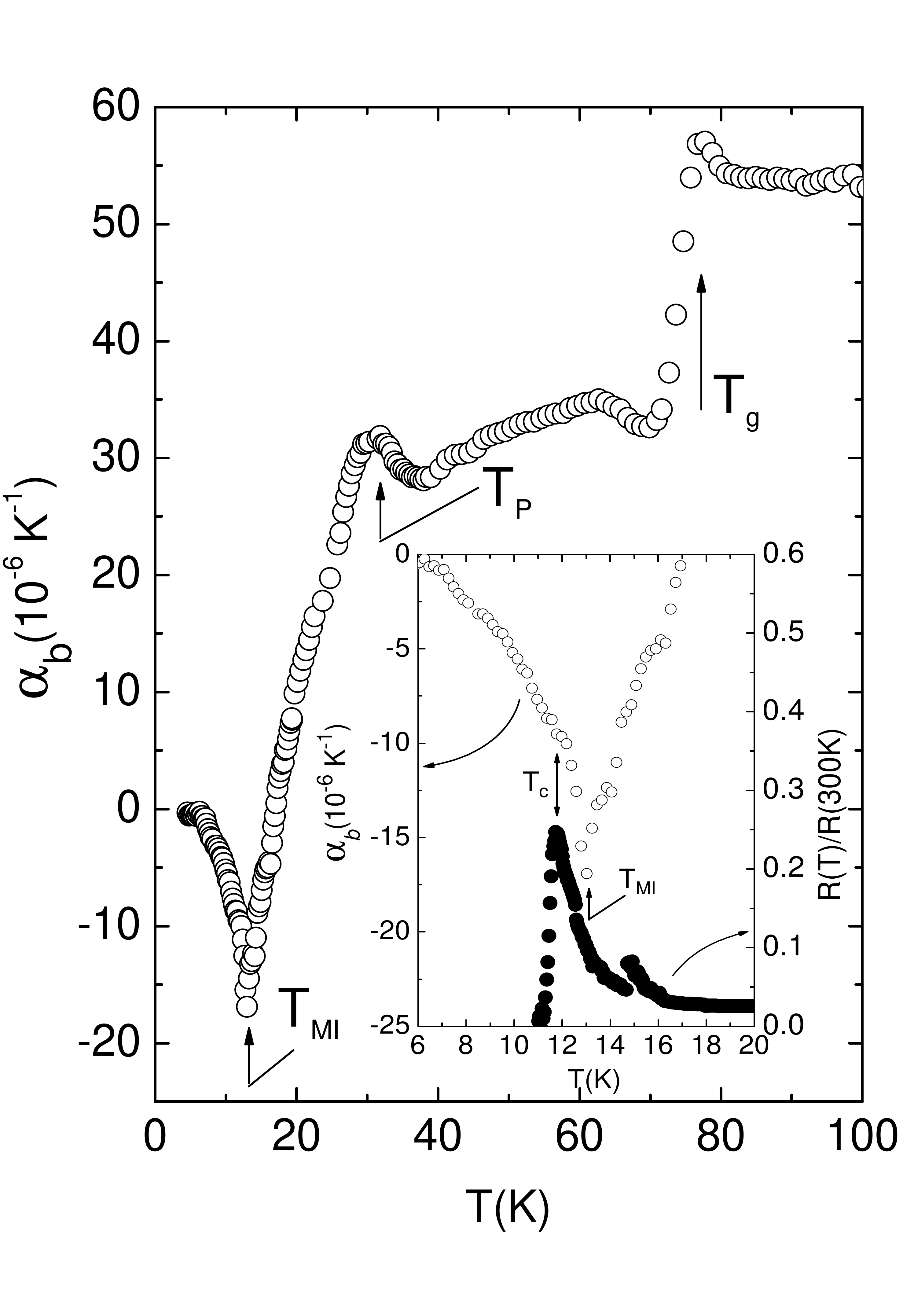}
\end{center}
\caption{Main panel: $\alpha(T)$ along the \emph{b}-axis for single crystalline $\kappa$-(D8-ET)$_{2}$Cu[N(CN)$_{2}$]Br. Inset:
blow-up of the low-temperature $\alpha_b(T)$ data (left scale) together with results of the
resistance normalized to the room temperature value (right scale).
$T_{MI}$ refers to the MI transition temperature and $T_c$ denotes
the critical temperature to percolative SC.}\label{Fig.2}
\end{figure}

The thermal expansion coefficient along the interplane \emph{b}-axis
in zero magnetic field is displayed in Fig.\,\ref{Fig.2}. Upon
cooling, an anomaly at $T_g$ $\approx$ 77\,K is observed. This
anomaly has been attributed to a glasslike transition, which has
been discussed in the literature in connection with the freezing out
of the ethylene end groups\cite{Mueller 02} and
anion-ordering\cite{Wolter, PhD}. Upon further cooling, a second
anomaly around $T_P$ = 30\,K is observed. This feature has been
assigned to critical fluctuations associated with the second-order
critical-end point of the first-order line \cite{Mariano, Bartosch},
indicated by the full up triangle in Fig.\,\ref{Fig.1}. Further
decreasing of the temperature reveals a pronounced anomaly around
$T_{MI}$ = 13.6\,K. As discussed in Ref.\cite{Mariano}, this
pronounced negative expansivity peak reflects the $b$-axis lattice
effect upon crossing the first-order MI transition line; cf.\,the inset of
Fig.\,\ref{Fig.2}, showing the anomaly in $\alpha_b(T)$ at $T$ =
$T_{MI}$ which coincides with the metal-insulator transition revealed by resistivity. The peak of the anomaly in $\alpha_b(T)$ was taken
as the thermodynamic transition temperature. The first-order
character of the transition was confirmed by the observation of
hysteresis in both $R$($T$) and relative length changes ($\Delta$\emph{l}/\emph{l}) around
$T_{MI}$ for another crystal
\cite{Mariano}. The drop of the resistance at $T_c$ = 11.6\,K,
accompanied by a tiny kink (indicated by the arrow) in
$\alpha_b(T)$, are signatures of percolative SC in minor portions of
the sample volume coexisting with the AFI state, see e.g.
\cite{Taniguchi,Jens}.

\begin{figure}
\begin{center}
\includegraphics[angle=0,width=0.40\textwidth]{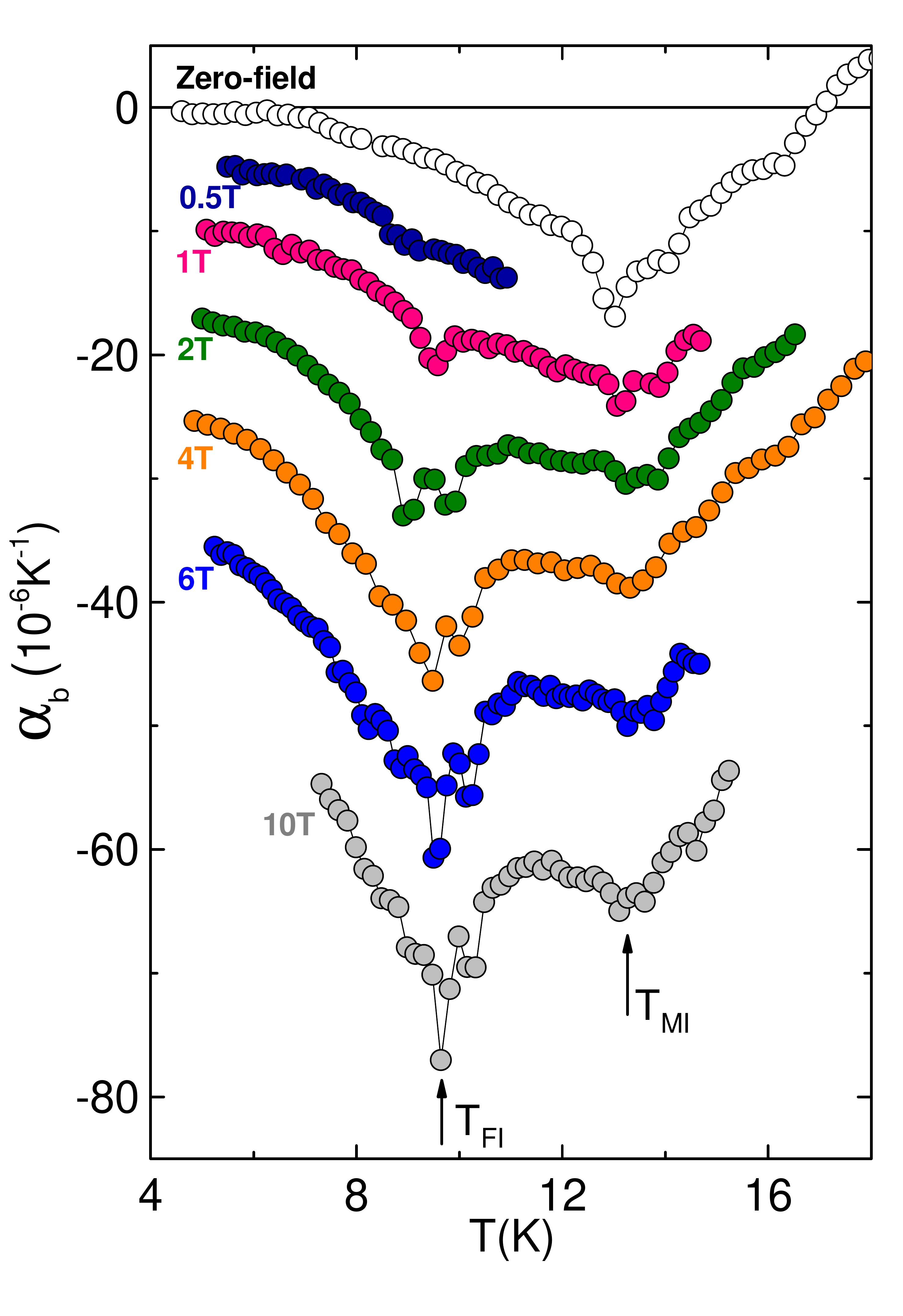}
\end{center}
\caption{(color online). $\alpha(T)$ for single crystalline
$\kappa$-(D8-ET)$_{2}$Cu[N(CN)$_{2}$]Br measured along the
\emph{b}-axis at low temperatures under selected fields, as
indicated. Solid lines are guide for the eyes. $T_{FI}$ refers to
the magnetic field-induced transition temperature and $T_{MI}$ to
the metal-to-insulator transition temperature. Data taken at 0.5, 1,
2, 4, 6 and 10\,T are shifted vertically for clarity.}\label{Fig.3}
\end{figure}

\begin{figure}
\begin{center}
\includegraphics[angle=0,width=0.44\textwidth]{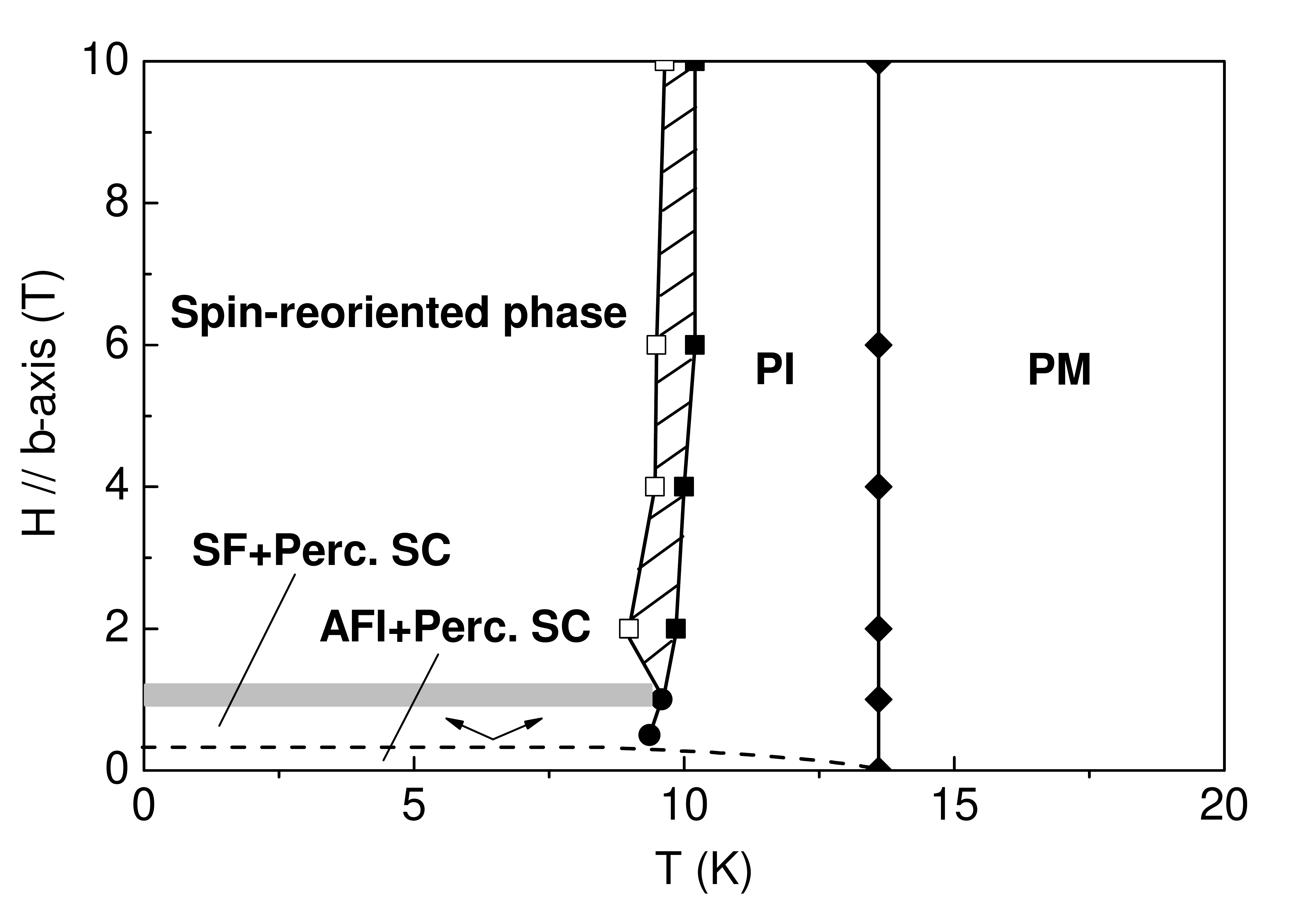}
\end{center}
\caption{Schematic \emph{$B$}-\emph{$T$} diagram for $\kappa$-D8-Br.
Symbols refer to the peak anomaly in $\alpha_b(T)$. Lines represent
the various phase transitions discussed in the main text. PM and PI
denote paramagnetic-metal and paramagnetic insulator, respectively.
Hatched region indicates $T$ window in which a double-peak structure
in $\alpha_b(T)$ is observed. Spin-reoriented phase refers to the
region where $\textbf{M}^\dagger_A$ and $\textbf{M}^\dagger_B$ are
antiparallel giving rise to an interplane antiferromagnetic ordering
as discussed in the main text.}\label{Fig.4}
\end{figure}

In what follows is a discussion on the effects of magnetic fields in the vicinity of
the Mott MI transition. Fig.\,\ref{Fig.3} shows the zero field
thermal expansion coefficient along the \emph{b}-axis at low
temperatures on expanded scales together with data taken in magnetic fields of 0.5, 1, 2, 4, 6 and 10\,T. The
first information obtained from these data sets is that upon
cooling under magnetic field up to 10\,T, $T_{MI}$ remains virtually
unaffected within the resolution of our experiment. Upon
further cooling under a weak field of 1\,T, however, a second
negative peak, centered around $T_{FI}$ = 9.5\,K, can be observed. This second peak
becomes more pronounced at 2\,T and saturates in size at a field
of about 4\,T. A closer inspection of the data in Fig.\,\ref{Fig.3} reveals
the existence of a double-peak structure for fields exceeding 1\,T.
It is well known that, under certain conditions, an antiferromagnet
exposed to a magnetic field can become unstable against a state
where the sublattice magnetization aligns approximately
perpendicular to the field $-$ the so-called spin-flop (SF) phase.
The SF transition occurs when the magnetic field is applied parallel
to the \emph{easy}-axis of the antiferromagnet and exceeds a
critical value (critical field). As we will discuss below in more
detail, the field dependence of $\alpha_b(T)$ in the present case is
indicative of a SF transition with strong coupling between the spin
and lattice degrees of freedom. As can be seen in Fig.\,\ref{Fig.3},
$H_c$ $<$ 1\,T ($H_c$ refers to the critical field) for
$\kappa$-D8-Br, which is consistent with a critical field $H_c$
$\sim$ 0.4\,T for the $\kappa$-Cl salt deduced in
Ref.\,\cite{Miyagawa}. However, the growth of the anomaly in
$\alpha_b$(\emph{T}) with increasing fields $H$ $>$ $H_c$ at a
virtually constant transition temperature $T_{FI}$ = 9.5\,K, is not
expected for a SF transition and requires an explanation.
In this respect we mention the results of resistance measurements on
50\% and 75\% deuterated $\kappa$-phase (ET)${_2}$Cu[N(CN)$_{2}$]Br
samples performed under field sweep (field applied along the
\emph{b}-axis) at $T$ = 5.50\,K and 4.15\,K, respectively, where a
transition from SC to a low-resistive state was found for fields
exceeding about 1\,T \cite{Taniguchi1}. This state was found to
transform into a high-resistive state via jump-like increases in the
resistance upon further increasing the field to 10\,T
\cite{Taniguchi1}. This behavior was interpreted by the authors as a
field-induced first-order SC-to-Insulator transition and related to
the theoretical scheme based on the SO(5) symmetry for
superconductivity and antiferromagnetism, first proposed by Zhang
\cite{Zhang} for the high-$T_c$ cuprates. Likewise, a drastic
$B$-induced increase in the resistance was observed also for the
$\kappa$-(ET)${_2}$Cu[N(CN)$_{2}$]Cl system when tuned close to the
MI transition by hydrostatic pressure \cite{Kagawa 04}. Here it was
argued that the marginally metallic/superconducting phases near the
Mott transition undergo a field-induced localization transition in
accordance with theoretical predictions, see, e.g.,
Ref.\,\cite{Georges 96} and references cited therein. Hence, based
on these observations we expect that close to the Mott-MI transition
in the present fully deuterated $\kappa$-(ET)${_2}$Cu[N(CN)$_{2}$]Br
salt the electronic states may undergo drastic changes upon
increasing the field. This may explain the growth of the anomaly in
$\alpha_b$(\emph{T}) with increasing fields $H$ $>$ $H_c$. In fact,
the initial rapid growth and the tendency to saturation above about
4\,T is very similar to the evolution of the resistance, i.e.\, the increase of $R(B, T = const.)$ with field
revealed in the above-mentioned transport studies
\cite{Taniguchi1,Kagawa 04}.
The observation of a field-induced transition at $T_{FI}$ $<$
$T_{MI}$ raises the question on the nature of the intermediary phase
that forms in the temperature window between $T_{MI}$ and $T_{FI}$
for fields $H$ $>$ $H_c$. In the present stage of the
investigations, i.e.\,by lacking a detailed magnetic characterization
of this state, we speculate that this phase represents a
paramagnetic insulator (PI), consistent with the notion of the SF
transition, cf.\,the preliminary phase diagram shown in
Fig.\,\ref{Fig.4}.
We stress that no such field-induced effects were observed for fields along the
\emph{a}- and \emph{c}-axis (not shown), also corroborating our claim of a SF transition. In fact, in
Ref.,\cite{Dylan1} the authors pointed out that the nature of the
interlayer magnetic ordering depends on the direction of the applied
magnetic field. In particular, based on a detailed analysis of NMR
and magnetization data, taking into account the
Dzialoshinskii-Moriya interaction, they found that antiferromagnetic
ordering between planes can be observed \emph{only} for magnetic
fields above $H_c$ applied along the \emph{b}-axis. At this point,
it is worth mentioning that the term \emph{"interplane afm ordering"}
is used here as defined in Figs.\,4 and 5 of Ref.\,\cite{Dylan1}.
Hence, given the absence of lattice effects at $T_{FI}$ for magnetic
fields applied along the \emph{a}- and \emph{c}-axis, our findings
indicate a close relation between interplane antiferromagnetic
ordering and the lattice effects observed at $T_{FI}$. Following the
notation used by the authors in Ref.\,\cite{Dylan1}, the
magnetization of the $+$($-$) sublattice at the layer $l$ is
$\textbf{M}_{+(-)l}$. The staggered and ferromagnetic moments are
given by $\textbf{M}^\dagger_l$ = ($\textbf{M}_{+l}$ $-$
$\textbf{M}_{-l}$)/2 and $\textbf{M}^F_l$ = ($\textbf{M}_{+l}$ $+$
$\textbf{M}_{-l}$)/2, respectively. For fields above $H_c$ applied
along the $b$-axis, $\textbf{M}^F_l$ is along $b$ and
$\textbf{M}^\dagger_l$ is in the $a$-$c$ plane.
$\textbf{M}^\dagger_A$ and $\textbf{M}^\dagger_B$ are antiparallel
giving rise to an interplane antiferromagnetic ordering. Our observation of a negative peak
anomaly in $\alpha_b(T)$ at  $T$ = $T_{FI}$ thus suggests that in
order to achieve this particular spin configuration, the gain in exchange energy forces the layers formed by the
(ET)$^+_2$ dimers to move apart from each other.
Similar dilatometric experiments were carried out on two other $\kappa$-D8-Br
single crystals. For both crystals, in contrast to the sharp anomaly
at $T_{FI}$ $\approx$ 9.5\,K under magnetic fields shown in
Fig.\,\ref{Fig.3}, the effects of magnetic fields result in a smooth
change of $\alpha_b(T)$
around the same temperature. Two factors should be considered as a
possible explanation for this: i) the crystal alignment upon
mounting the sample in the dilatometer. As reported in the
literature, see e.g. \cite{Pankhurst}, SF transitions are strongly
dependent on the direction of the applied magnetic field. A minute
misalignment of the material's \emph{easy}-axis with respect to the applied magnetic field can
cause a suppression of the transition; ii) sample inhomogeneities
and/or the effect of the pressure exerted by the dilatometer on the
sample.
For one of these single crystals, we studied the effect of
quasi-uniaxial pressure exerted by the dilatometer on the sample and
observed that a quasi-uniaxial pressure of some tenth of bars is
enough to change the shape of the thermal expansion curves
considerably: upon the application of uniaxial pressure of about 65\,bar the transition was smeared out over a very wide
temperature range \cite{PhD}. Our thermal expansion results at magnetic fields
along the \emph{b}-axis are summarized in the schematic $H$ versus
$T$ diagram depicted in Fig.\,\ref{Fig.4}. The dashed line around
$\sim$ 0.5\,T separates the AFI from the SF phase, while the thick
line marks the suppression of percolative SC together with the
appearance of the spin-reoriented phase.

\section{Summary}
In conclusion, our thermal expansion studies on fully deuterated
$\kappa$-(BEDT-TTF)$_2$Cu[N(CN)$_{2}$]Br under magnetic field reveal
the insensitivity of the Mott metal-to-insulator transition
temperature under fields up to 10\,T, which is in accordance with
the proposal of a Mott insulating state with a $\pi$-hole localized
in a dimer. A field-induced phase transition at $T_{FI}$ = (9.5
$\pm$ 0.5)\,K is observed, indicative of a spin reorientation with
strong magneto-elastic coupling.
Further experiments, including magnetostriction measurements both
below and above $T_{FI}$ as well as magnetic measurements with $B$
thoroughly aligned along the $b$-axis, will help to better
understand the behavior of almost localized strongly correlated
electrons in this interesting region of the phase diagram of the
$\kappa$-phase (BEDT-TTF)$_2$X charge-transfer salts.  A theoretical
study on the stability of the Mott insulating state and the adjacent
superconducting phase, taking into account field-induced lattice
effects, is highly desired.\newline

\small M. de S. acknowledges financial support from the S\~ao Paulo
Research Foundation - Fapesp (Grants 2011/22050-4, 2012/02747-3).

\end{document}